\documentclass{article}
\title{A Proposed Alternative Low Energy Quantum Field Theory of Gravity Based on a
Bose-Einstein Condensate Effect}
\author{Alexander Oshmyansky}
\begin{document}
\maketitle

\begin{abstract}
An alternative quantum field theory for gravity is proposed for
low energies based on an attractive effect between contaminants in
a Bose-Einstein Condensate rather than on particle exchange. In
the ``contaminant in condensate effect," contaminants cause a
potential in an otherwise uniform condensate, forcing the
condensate between two contaminants to a higher energy state. The
energy of the system decreases as the contaminants come closer
together, causing an attractive force between contaminants. It is
proposed that mass-energy may have a similar effect on Einstein's
space-time field, and gravity is quantized by the same method by
which the contaminant in condensate effect is quantized. The
resulting theory is finite and, if a physical condensate is
assumed to underly the system, predictive. However, the proposed
theory has several flaws at high energies and is thus limited to
low energies. Falsifiable predictions are given for the case that
the Higgs condensate is assumed to be the condensate underlying
gravity.
\end{abstract}

\section{Proposed Theory on a Scalar Field}

Within Bose-Einstein Condensates \cite{Pethick02}, the authors
predict in a separate paper \cite{Oshmyansky07} the existence of a
new effect which causes an attractive force between two
contaminants, the ``contaminant in condensate" (CIC) effect. It is
proposed that contaminants act as a potential within the
condensate. This causes the condensate in between two contaminates
to jump to a higher energy state than if no contaminants existed.
By assuming that the condensate behaves as a massive scalar field
governed by:

\begin{equation}
\frac{1}{c^2}\frac{\partial^2\varphi(t,x)}{\partial
t^2}-\frac{\partial^2\varphi(t,x)}{\partial x^2} +
\frac{m^2c^2}{\hbar^2}\varphi(t,x)=0
\end{equation}

\noindent with induced standing waves between contaminants
governing the condensate superstate given by:

\begin{equation}
\varphi^{(\pm)}_n(t,x) = \sqrt{\frac{c}{a \omega_n}} e^{\pm i
\omega_n t} \sin k_nx
\end{equation}

\begin{equation}
 \omega_n = \sqrt{\frac{m^2c^4}{\hbar^2}+c^2k_n^2}
\end{equation}

\begin{equation}
 k_n=\frac{\pi n}{a}, n = 1,2,...,
\end{equation}

\noindent the expectation value of energy associated with the
superstate over all energy levels is determined to be:

\begin{equation}
E(a) \approx -\frac{mc^2}{4}-\frac{\pi\hbar c}{24a} + \frac{\hbar
c}{23\pi a}\mu^2 \ln \mu
\end{equation}

\noindent with $\mu \equiv mca/\hbar$ in the case $\mu \ll 1$.
Note for a massless field this becomes:

\begin{equation}
E(a) = -\frac{\pi\hbar c}{24a}.
\end{equation}

\noindent These results were derived by the same method by which
the Casimir effect is derived \cite{Bordag06}.

However, in a physical Bose-Einstein condensate, energy levels are
so low that, as argued in \cite{Oshmyansky07}, induced superstates
are likely always in their lowest energy state available. In order
to make a more accurate model of the force associated with the CIC
effect, one must therefore find the energy associated with the
creation of a superstate and its change to a different size. To
simplify both the scattering calculations and the creation of an
S-matrix to describe the CIC effect, the approach taken to
determining the energy of an induced superstate is to associate a
scalar particle propagator with the condensate superstate:

\begin{equation}
\frac{1}{k^2-m^2+i\varepsilon}.
\end{equation}

\noindent Again, as all energy states are not integrated over as
in the Casimir effect, it is safe to manipulate one state at a
time in calculations.

The superstate ``propagates" in the space of distances rather than
physical space however. That is, a superstate is said to
``propagate" from one distance to another, as describing a
condensate superstate by a single point would incompletely
describe its position. Taking into account this philosophical
point, the standard machinery of QFT is used. A force between two
particles is then produced by the creation of a superstate and its
movement. For a massless field, this results in a force:

\begin{equation}
F=-\frac{1}{4\pi a^2}
\end{equation}

\noindent as usual \cite{Zee03} \cite{Kaku93}.

Note two particles can occupy the same point in distance-space,
preventing superstate interaction, and the superstate does not
need to interact with any particles to create a force. As this
occurs in the CIC effect, it thus fulfills two physical
requirements that are required of a QFT of the CIC effect.

Also it should be noted that with no interactions among particles
allowed, the Feynman rules for our theory are trivial.

\section{A Physical Condensate Underlying Gravity?}

A finite quantum field theory has thus been defined, essentially
by fiat. All interactions which could cause a divergence have been
eliminated as only creation, propagation, and annihilation are
allowed. The model can further be extended to a spin-2, massless
tensor field to find a finite, though not predictive, quantum
theory of gravity, as will be demonstrated below.

In condensates, though, only energy outside of the condensate will
serve as a potential. Thus if a physical condensate is used as the
source of gravity in our quantum field theory, higher order
self-interaction terms can be ignored as unphysical. A predictive
theory of gravity can thus be created. (Note that if the Higgs
condensate is assumed to underly gravity as will be the case in
section (3), this model would explain why the Higgs condensate has
no apparent ``weight" and its energy density is not observed, a
problem noted in \cite{Wilcsek05}.)

In order to preserve relativity, all particles interacting through
a condensate must be separated by either a time-like distance or
light-like distance. Also, the energy for the creation of a
superstate of many particles comes from each individual particle,
lowering the temperature of the system as a whole.

The resulting theory starts with the graviton propagator in the
harmonic gauge as usual \cite{Zee03}, but redefines it in distance
space so it can apply to a particle superstate rather than a
particle. This gives:

\begin{equation}
D_{\mu\nu,\lambda\sigma}(k)=\frac{1}{2}\frac{\eta_{\mu\lambda}\eta_{\nu\sigma}+\eta_{\mu\sigma}\eta_{\nu\lambda}
- \eta_{\mu\nu}\eta{\lambda\sigma}}{k^2+i\varepsilon}
\end{equation}

\noindent where $\eta$ is the Minkowski metric and
$g_{\mu\nu}=\eta_{\mu\nu} + h_{\mu\nu}$ where $h_{\mu\nu}$ are
deviations from the Minkowski metric. This couples to the
stress-energy tensor $T^{\mu\nu}$ defined according to the
variation of the matter action $S_M$ by:

\begin{equation}
T^{\mu\nu}(x)=-\frac{2}{\sqrt{-g}}\frac{\delta S_M}{\delta
g_{\mu\nu}(x)}
\end{equation}

\noindent and gives the scattering amplitude:

\begin{equation}
GT^{\mu\nu}_{(1)}D_{\mu\nu,\lambda\sigma}(k)T^{\lambda\sigma}_{(2)}
= \frac{G}{2k^2}(2T^{\mu\nu}_{(1)}T_{(2)\mu\nu}-T_{(1)}T_{(2)}).
\end{equation}

\noindent Between non-relativistic matter, this becomes
$\frac{G}{2k^2}T^{00}_{(1)}T^{00}_{(2)}$. As usual, the Fourier
transform gives the interaction potential:

\begin{equation}
G\int\int d^3xd^3x'T^{(1)00}(x)T^{(2)00}(x')\int
d^3ke^{i\overrightarrow{k}\cdot(\overrightarrow{x}-\overrightarrow{x}')}
\frac{1}{\overrightarrow{k}^2}
\end{equation}

\noindent where a change is made from the graviton-exchange model
and one integrates over distances rather than positions. This
reduces to the Newtonian potential $\frac{GM_1M_2}{r}$.

This is an identical result to a graviton-exchange model
\cite{Zee03}, but interactions which cause divergences are not
predicted.

It should be noted that the above model bears resemblance to
Sakharov's ``Induced Gravity" model \cite{Sakharov68}, as both
speculate gravity to arise from underlying quantum fluctuations
rather than as a fundamental force. However, the mechanism by
which this is thought to occur is different in the model above.
Vacuum energy is not presumed as a basis for gravity. Rather,
superstates of a physical condensate mediate the gravitational
interaction.

\section{Physical Predictions}

In order for a theory of gravity to exist which works by the
massless mechanism in section (2), there must exist a field
associated with a spin-2, massless, tensor particle that energy
forms a potential in (perhaps associated with the Einstein's space
time field). A condensate of these particles would then be
sufficient to cause a gravity-like interaction. The theory thus
requires and therefore predicts the existence of some form of
massless condensate in which energy forms a potential.

The theory would also predict that there is no self-interaction
correction to the strength of the gravitational interaction.
However, this prediction would only occur at energies currently
not testable.

Alternatively, a condensate associated with gravity consisting of
massive particles, such as the Higgs condensate, would produce
testable effects. Note that there is nothing preventing a
condensate of massive particles from producing an effect such as
that described above which could be associated with gravity. This
is because the information about a superstate would still travel
at the speed of light. There are two chief effects predicted in
this case, which are described below.

It can be heuristically argued that the CIC effect in a condensate
composed of a massive field should behave no differently than in a
condensate composed of a massless field. This is because of the
argument that information about the condensate is massless, and
thus the CIC effect would still behave as though it were occurring
in a massless medium. The results below disregard this argument
and strictly follow the mathematics of the proposal.

\subsection{A Universal Repulsive Force and its role in Early Universe Cosmology}

As described in section (1), the energy caused by two contaminants
in a massive, scalar condensate is equal to, if we sum over all
excitation modes of a potential supersate:

\begin{equation}
E(a) \approx -\frac{mc^2}{4}-\frac{\pi\hbar c}{24a} + \frac{\hbar
c}{23\pi a}\mu^2 \ln \mu.
\end{equation}

\noindent This results in a force including arbitrary constants
$b_n$:

\begin{equation}
F(a) \approx -\frac{b_G}{a^2} - b_1 \ln b_2 a - b_3.
\end{equation}

In the early universe, uniform extreme high energy conditions
could potentially cause induced superstates to obtain higher
energies. It is thus proposed that the repulsive component of
equation (14) may play a role in inflationary cosmology. If the
Higgs condensate induces a gravity-like effect in the CIC
mechanism, then the following inflationary potential is predicted
to have occurred after symmetry breaking:

\begin{equation}
V(\phi) = b_1 \phi \ln (b_2 \phi) + b_3.
\end{equation}

This results in ``slow roll" parameters \cite{Tsujikawa03}:

\begin{equation}
\epsilon = \frac{m^2_{pl}}{16\pi}\cdot \frac{V'(\phi)}{V(\phi)} =
\frac{m^2_{pl}}{16\pi}\cdot \frac{b_1 \ln (b_2 \phi) + b_1}{\phi
\cdot b_1 \ln (b_2 \phi) + b_3}
\end{equation}

\begin{equation}
\eta = \frac{m^2_{pl}}{8\pi}\cdot \frac{V''(\phi)}{V(\phi)} =
\frac{m^2_{pl}}{8\pi}\cdot \frac{b_1}{\phi^2 \cdot b_1 \ln (b_2
\phi) + b_3}
\end{equation}

\noindent and a predicted number of e-foldings \cite{Freese04}:

\begin{equation}
N = \int_{t_i}^{t_f} H dt = \int_{\phi_i}^{\phi_f}
\frac{V(\phi)}{V'(\phi)} d\phi = \int_{\phi_i}^{\phi_f} \frac{\phi
\cdot b_1 \ln (b_2 \phi) + b_3}{b_1 \ln (b_2 \phi) + b_1} d\phi.
\end{equation}

\subsection{Universal Repulsive Force in the present day?}

As there is no reason for this potential to disappear after the
inflationary period (when the potential energy of the field
predominates over its kinetic energy) is over, there should thus
be a universal repulsive force between particles of order $O(\ln
a)$ presently. However, if the assumption that a condensate will
always be in its lowest energy state available is used, which is
perhaps more accurate, then this effect is not predicted. In fact,
the energy of massive condensate superstate creation and movement
is predicted by the path integral method to be $-\frac{1}{4\pi
a}e^{-ma}$ which is clearly a physically untenable potential.

It is still proposed however, that this potential force, with an
appropriately small coupling constant, could be a physical
justification for the apparent cosmological constant
\cite{Padmanabhan03}. However, for the proposed repulsive force to
be produced by this mechanism, the perhaps unphysical assumption
that a permeating condensate exists in arbitrarily high energy
states must be made. Also, unfortunately, we have found no
satisfactory method to incorporate this effect in Einstein's
equation as of the time that this is being written. This is a
major flaw in the theory of a massive condensate inducing gravity
and it may well be that there is no method of successfully
incorporating it into Einstein's equation. It is presently the
subject of ongoing work. It does, though, seem promising that some
form of potentially testable prediction can be made for certain
variations of a gravity as CIC effect theory.

\section{Conclusion}
We have attempted to show in as brief and straightforward a manner
possible that if there exists a field (such a spin-2 tensor field
or the Higgs field) associated with a particle that forms a
condensate which permeates space and in which mass-energy forms a
potential, a finite, predictive quantum field theory of gravity
can be developed by assuming gravity to be a CIC effect in the
condensate. A CIC effect in the Higgs condensate could produce a
finite, predictive quantum field theory of gravity with
falsifiable predictions, primarily a universal repulsive force of
order $O(\ln a$). However, there are serious problems with
incorporating the results of the predictions with Einstein's
equations.

Unfortunately, there are inherent problems with the CIC approach
approach at high energies, which is why it is only proposed as a
low energy theory. First, since predicted effects at very high,
early universe energies cannot be incorporated into Einstein's
equations at this time, it does not seem to be reducible to
Einstein's theory. Second, it is not background independent.

\section{Acknowledgements}
I would like to thank P.K. Maini as well as the rest of the
students and faculty at his lab group and at the Maths Institute
at Oxford. I would also like to thank Keith Burnett for his input
in the development of the initial Bose-Einstein condensate effect.
Funding for this project was provided by the Marshall Aid
Commemoration Commission.

\end{document}